\documentclass[useAMS, usenatbib, usegraphicx]{mn2e} 
\usepackage{amssymb}

\newcommand{\Mpch}{~h^{-1}~\mbox{Mpc}}

\newcommand{\Msunh}{~h^{-1}~\mbox{M}_{\odot}}

\newcommand{\K}{~\mbox{K}}

\newcommand{\eV}{~\mbox{eV}}
\newcommand{\erg}{~\mbox{erg}}
\newcommand{\kms}{~\mbox{km s}^{-1}}
\newcommand{\gadget}{{\sc gadget-2}}

\title[Star formation feedback amplification]{Photo-heating and supernova feedback amplify each other's effect on the cosmic star formation rate}

\author[Pawlik \& Schaye] 
       {Andreas H. Pawlik$^{1}$\thanks{E-mail: pawlik@strw.leidenuniv.nl}
	 and
	 Joop Schaye$^{1}$\thanks{E-mail: schaye@strw.leidenuniv.nl}\\
	 $^{1}$Leiden Observatory, Leiden University, P.O. Box 9513, 2300RA Leiden, The Netherlands}

\begin{document}

\date{Accepted; Received; in original form}

\pagerange{\pageref{firstpage}--\pageref{lastpage}} \pubyear{}

\maketitle

\label{firstpage}

\begin{abstract}
Photo-heating associated with reionisation and kinetic feedback from
core-collapse supernovae have previously been shown to suppress the
high-redshift cosmic star formation rate.  Here we investigate the
interplay between photo-heating and supernova feedback using a set of
cosmological, smoothed particle hydrodynamics simulations. We show
that photo-heating and supernova feedback mutually amplify each
other's ability to suppress the star formation rate. Our results
demonstrate the importance of the simultaneous, non-independent
inclusion of these two processes in models of galaxy formation to
estimate the strength of the total negative feedback they exert. They
may therefore be of particular relevance to semi-analytic models in
which the effects of photo-heating and supernova feedback are
implicitly assumed to act independently of each other.
\end{abstract}

\begin{keywords}
 galaxies: high-redshift -- galaxies: formation -- galaxies: evolution
\end{keywords}

\section{Introduction}
\label{Sec:Introduction}
The cosmic star formation rate (SFR) is an important observable of our
Universe. It is affected by a variety of physical processes, many of which
are in turn regulated by the SFR, giving rise to so-called feedback
loops (for an overview see, e.g., \citealp{Ciardi:2005}).  Photo-ionisation
heating due to the absorption of ionising photons from star-forming regions
and the injection of kinetic energy from supernova (SN) explosions of massive stars
provide two such feedback loops. Their implications for the assembly of the
first generation of galaxies have been extensively discussed in studies of the
epoch of reionisation (for a review of this epoch see, e.g., \citealp{Loeb:2001}). 
\par
Photo-heating associated with reionisation increases the mean temperature of the intergalactic medium (IGM) to
$\sim 10^4 \K$ (e.g., \citealp{Hui:1997}) and reduces the rate at
which hotter gas can cool (\citealp{Efstathiou:1992,Wiersma:2008}).
The increase in the gas temperature keeps
the IGM smooth and prevents the assembly 
of low-mass galaxies, that is, galaxies with masses corresponding to a virial
temperature $\lesssim 10^4 \K$ (e.g., \citealp{Shapiro:1994};
\citealp{Gnedin:1998}). Moreover, the gas in galaxies
that have already collapsed is relatively quickly photo-evaporated (e.g.,
\citealp{Shapiro:2004}; \citealp{Iliev:2005}), strongly decreasing the gas
fraction of low-mass halos (e.g., \citealp{Thoul:1996};
\citealp{Barkana:1999}; \citealp{Dijkstra:2004}; \citealp{Susa:2006}). 
Indeed, the cosmic SFR has been predicted to exhibit a
distinct drop around the redshift of reionisation (\citealp{Barkana:2000}).
Photo-ionisation heating is therefore said to provide a negative feedback on
reionisation.\footnote{As pointed out by \cite{Pawlik:2008},
photo-heating also provides a strong positive feedback on reionisation because
the increase in the gas temperature smoothes out density fluctuations,
reducing the recombination rate.}
\par
SN explosions of massive stars typically inject a few solar masses of
gas with velocity of $\sim 10^4\kms$, corresponding to a kinetic energy of
$\sim 10^{51} \erg$. The ejected material sweeps up and shock-heats the
surrounding gas, entraining outflows sufficiently powerful to, at least
temporarily, substantially reduce the gas fractions for galaxy-scale dark
matter halos (e.g., \citealp{Yepes:1997}; \citealp{Scannapieco:2006}). 
Since this leads to a suppression of the SFR, SN explosions, like
photo-heating from reionisation, provide a negative feedback on
reionisation. In addition to the depth of the gravitational potential,
the ability of SN feedback to reduce the gas fractions generally depends on the
geometry of the gas distribution (e.g., \citealp{MacLow:1999}). 
\par
Studies of the effects of photo-heating and SN feedback on the SFR
that considered each process in isolation have been augmented by
studies that included both photo-heating and SN feedback. These
studies include simulations of the formation of the first stars (e.g.,
\citealp{Greif:2007}; \citealp{Wise:2008}; \citealp{Whalen:2008}), of
the evolution of isolated galaxies (e.g., \citealp{Fujita:2004};
\citealp{caius:2008}, \citealp{Tasker:2008}), and of galaxies in a cosmological volume
(e.g., \citealp{Tassis:2003}). Some of these studies also investigate
the interplay between photo-heating and SN feedback. For example,
\cite{Kitayama:2005} demonstrated in a one-dimensional hydrodynamical
study that a previous episode of photo-heating may increase the
efficiency of the evacuation of dark matter halos by (thermal) SN
feedback and enable the destruction of galaxies out to much larger
masses. In this letter we report on another interaction of 
star formation feedbacks, using three-dimensional galaxy formation simulations.
\par
We employ a set of Smoothed Particle Hydrodynamics (SPH)
cosmological simulations that include star formation, and photo-ionisation
heating from a uniform ultraviolet (UV) background and/or kinetic feedback
from core-collapse SNe. We investigate how photo-heating
affects the high-redshift ($z \ge 6$) SFR in the presence and absence
of SN feedback and, conversely, how SN feedback affects the
cosmic SFR in the presence and absence of a photo-ionising background. We find
that the inclusion of SN feedback amplifies the suppression of the
cosmic SFR due to the inclusion of photo-heating.  On the other hand, the inclusion
of photo-heating amplifies the suppression of the cosmic SFR due to the
inclusion of SN feedback.
\par
Photo-heating and SN feedback therefore mutually amplify each other in
suppressing the SFR. Our results are relevant to current implementations of 
(semi-) analytic models of galaxy formation (see \citealp{Baugh:2006} for a review),
in which the effects of photo-heating and SN feedback are implicitly assumed to act independently
of each other. These models may thus 
underestimate the strength of the combined negative feedback from photo-heating and SNe.
\par
We emphasise that none of our simulations has a
sufficiently high resolution to achieve convergence in the cosmic SFR. Future
simulations will therefore be required to quantify our qualitative
statement. We show, however, that the factor by which photo-heating 
and SN feedback amplify each other's ability to suppress the cosmic SFR becomes larger
with increasing resolution, giving credibility to our main conclusion.
\par
This letter is organised as follows. We
present our simulation method in \S\ref{Sec:Simulations} and we
illustrate our 
main result in \S\ref{Sec:Results}. Finally, we summarize our
conclusions and discuss the caveats inherent to the present work in
\S\ref{Sec:Discussion}.  

\section{Simulations}
\label{Sec:Simulations}
\begin{table}
\begin{center}
\caption{Simulation parameters. From left to right, table entries are:
  simulation label;
  comoving size of the simulation box, $L_{\rm box}$;
  number of DM particles, $N_{\rm dm}$; mass of dark matter particles, 
  $m_{\rm dm}$. A prefix \textit{r9} indicates the
  inclusion of photo-heating in the optically thin limit from a uniform UV background (for $z \le 9$) and a suffix \textit{winds}
  indicates the inclusion of SN feedback (with initial wind velocity $v_{\rm w} = 600 \kms$
  and mass loading $\eta = 2$). A bold font marks our set of reference simulations. 
\label{tbl:params}}
\begin{tabular}{cccccc}

\hline
\hline
simulation &$L_{\rm box}$ & $N_{\rm dm}$ & $m_{\rm dm}$   & \\
        & $[\Mpch]$   &             & $[10^5\Msunh]$           &\\           
\hline

\bf \textit{L6N256}        & $ \bf 6.25$ & $ \bf 256^3$ & $ \bf 8.6$  &\\
\bf\textit{r9L6N256}        & $ \bf 6.25$ & $\bf 256^3$ & $\bf 8.6$  &\\
\bf\textit{L6N256winds}        & $\bf 6.25$ & $\bf 256^3$ & $\bf 8.6$  &\\
\bf\textit{r9L6N256winds}        & $\bf 6.25$ & $\bf 256^3$ & $\bf 8.6$ &\\
\textit{L3N256}        & $ 3.125$ & $ 256^3$ & $ 1.1$  &  \\
\textit{r9L3N256}        & $ 3.125$ & $ 256^3$ & $ 1.1$  &\\
\textit{L3N256winds}        & $ 3.125$ & $ 256^3$ & $ 1.1$ &\\
\textit{r9L3N256winds}        & $ 3.125$ & $ 256^3$ & $ 1.1$ &\\
\textit{L3N128}        & $ 3.125$ & $ 128^3$ & $ 8.6$  & \\
\textit{r9L3N128}        & $ 3.125$ & $ 128^3$ & $ 8.6$  &\\
\textit{L3N128winds}        & $ 3.125$ & $ 128^3$ & $ 8.6$  &\\
\textit{r9L3N128winds}        & $ 3.125$ & $ 128^3$ & $ 8.6$ &\\

\hline
\end{tabular}
\end{center}
\end{table}
\par
Our simulation method is identical to that employed in \cite{Pawlik:2008}, to
which we refer the reader for more details. We use a modified version of the
N-body/TreePM/SPH code \gadget\ (\citealp{Springel:2005}) to perform a total
of 12 cosmological SPH simulations at different resolutions, using different
box sizes. We employ the set of cosmological parameters $[\Omega_{\rm{m}},
\Omega_{\rm{b}}, \Omega_\Lambda, \sigma_8, n_{\rm{s}}, h]$ given by $[0.258,
0.0441, 0.742, 0.796, 0.963, 0.719]$, in agreement with the WMAP 5-year
observations (\citealp{Komatsu:2008}).  The simulations include radiative
cooling, star formation and, optionally, photo-heating by a uniform UV
background and/or kinetic feedback from SNe (see
Table~\ref{tbl:params}).
\par
The gas is of primordial composition and is 
allowed to cool by collisional ionisation and excitation, emission of
free-free and recombination radiation and Compton cooling off the cosmic
microwave background.  Molecular hydrogen is kept photo-dissociated at all
times by the inclusion of a soft UV background. We employ the star formation 
recipe of \cite{Schaye:2008}, to which we refer the reader for details.  According to this recipe, 
gas forms stars at a pressure-dependent rate that reproduces the observed
Kennicutt-Schmidt law (\citealp{Kennicutt:1998}).
\par
Photo-ionisation (heating) is included in the optically thin limit using a
uniform \cite{Haardt:2001} UV background from quasars and galaxies for
redshifts $z \le z_{\rm r} = 9$.  The value for $z_{\rm{r}}$ is consistent
with the most recent determination of the Thomson optical depth towards
reionisation from the WMAP (5-year) experiment (\citealp{Komatsu:2008}).  We
inject an additional thermal energy of $2 \eV$ per proton at $z = z_{\rm{r}}$.  In the
absence of shocks, gas particles at the cosmic mean density are therefore kept
at a temperature $T \approx 10^4 \K$ for $z<z_{\rm r}$ (see Fig.~1
of \citealp{Pawlik:2008}).
\par
We model kinetic feedback from star formation using the prescription of
\cite{caius:2008}, according to which core-collapse SNe locally
inject kinetic energy and kick gas particles into winds. The feedback is
specified by two parameters, the initial wind mass loading in units of the
newly formed stellar mass, $\eta$, and the initial wind
velocity $v_w$. We adopt $\eta = 2$ and $v_w = 600 \kms$, consistent with
observations of local (e.g.~\citealp{Veilleux:2005}) and redshift $z \approx
3$ (e.g.~\citealp{Shapley:2003}) starburst galaxies. This choice of
parameters implies that 40 per cent of the energy available from
core-collapse SNe is injected as kinetic energy (assuming a Chabrier
initial mass function in the range $0.1 \le M/M_\odot \le 100$ and
$10^{51}\,{\rm erg}$ per $M> 6~M_\odot$ star), while the remaining 60 per cent 
are implicitly assumed to be lost radiatively.
\par
We use a Friends-of-Friends halo finder with linking length $b = 0.2$ to
obtain a list of dark matter halos contained in each of our simulation outputs. Only
halos containing more than 100 dark matter particles are included in these
lists.  For each simulation output we compute the SFR associated with 
dark matter halo as follows. First, gas particles are attached to the nearest dark matter particle. Second,
the SFR of a dark matter halo is the sum of the SFRs of the gas particles that
were attached to the dark matter particles it contains.
\par
\begin{figure}
  \includegraphics[width=0.49\textwidth]{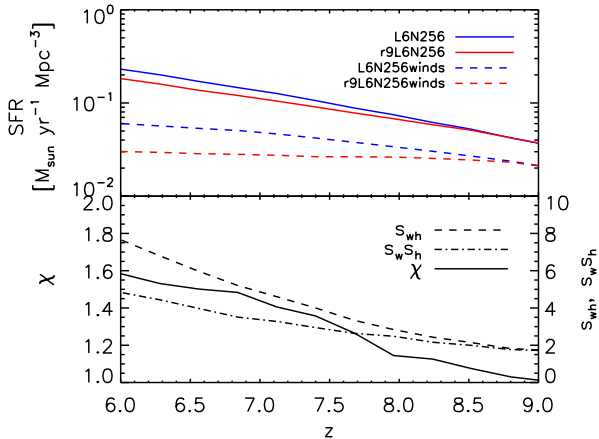}
  \caption{Top panel: evolution of the total SFRs in our set of
    reference simulations. Photo-heating and SN feedback were included
    for the red and dashed curves, respectively. Star formation is strongly suppressed when
    photo-heating and/or kinetic feedback from SNe are
    included.  Bottom panel: Suppression factors $s_{\rm wh}$ and
    $s_{\rm w}s_{\rm h}$ (right y-axis) and feedback amplification factor $\chi
    \equiv s_{\rm wh}/(s_{\rm w}s_{\rm h})$ (left y-axis). The fact that
    $\chi>1$ implies that photo-heating and SN feedback amplify each
    other's ability to suppress the SFR.}
  \label{Fig:1}
\end{figure}

\section{Results}
\label{Sec:Results}
Our main result is shown in Fig.~\ref{Fig:1}. The top panel shows the
evolution of the total SFR, defined as the sum of the star formation rates
over all halos, for our set of reference simulations. We use $\dot{\rho}_{*, \rm wh}$, $\dot{\rho}_{*,
\rm h}$ and $\dot{\rho}_{*, \rm w}$ to denote, respectively,
the SFR densities in the simulations with both SN feedback and photo-heating (red dashed curve,
\textit{r9L6N256winds}), with photo-heating but without SN
feedback (red solid curve, \textit{r9L6N256}) and with SN
feedback but without photo-heating (blue dashed curve, \textit{L6N256winds}).
We denote the SFR density in the simulation that included
neither photo-heating nor wind feedback with $\dot{\rho}_{*}$ (blue
solid curve, \textit{L6N256}). 
\par
Both the inclusion of photo-heating and the inclusion of kinetic feedback from
SNe lead to a significant suppression of the SFR. The factor $s_{\rm h} \equiv
\dot{\rho}_{*}/ \dot{\rho}_{*,\rm h}$ by which the SFR is
suppressed due to photo-heating is smaller than the factor $s_{\rm w}
\equiv \dot{\rho}_{*}/ \dot{\rho}_{*,\rm w}$ by which the SFR is
suppressed due to kinetic feedback. As expected, the simultaneous inclusion of
photo-heating and feedback from SNe leads to a suppression of the SFR
by a factor $s_{\rm wh} \equiv \dot{\rho}_{*}/ \dot{\rho}_{*,\rm wh}$ that is larger
than the factors $s_{\rm h}$ and $s_{\rm w}$ by which the SFRs are suppressed due to
the sole inclusion of either photo-heating or SN feedback.
\par
Interestingly, the factor by which the SFR is suppressed due to photo-heating
is larger in the presence (set of dashed curves) than in the absence (set of
solid curves) of SN feedback. Conversely, the factor by which the SFR
is suppressed due to SN feedback is larger in the presence (set of red
curves) than in the absence (set of blue curves) of
photo-heating. 
\par
Photo-heating and SN feedback thus mutually amplify each
other in suppressing the SFR. This amplification probably arises because
the inclusion of photo-heating keeps the gas diffuse, which makes it easier for the winds 
to drag it out of halos. Winds, on the other hand, move gas from 
the central to the outer parts of halos, where it is more susceptible to the 
photo-evaporation process. Models that implicitly ignore this interaction between 
photo-heating and SN feedback, like for example (semi-) analytic models of galaxy formation (e.g., \citealp{Khochfar:2008};
\citealp{Monaco:2007}; \citealp{Benson:2006}; \citealp{Croton:2006}; \citealp{Somerville:2002}),
thus may underestimate the strength of the feedback these processes exert. 
\par
We now define the feedback amplification factor $\chi
\equiv s_{\rm wh}/ (s_{\rm w} s_{\rm h})$.  A value $\chi = 1$ would indicate that
photo-heating and SN feedback suppress the SFR independently of each
other. A value $\chi > 1$ ($\chi < 1$)
would indicate that photo-heating and SN feedback amplify
(weaken) each other's ability to suppress the SFR. The evolution of
the amplification factor $\chi$ is shown in the bottom panel of
Fig.~\ref{Fig:1}, together with that of $s_{\rm wh}$ and $s_{\rm w} s_{\rm h}$. The fact that
$\chi > 1$ for $z<9$ implies that photo-heating and SN
feedback amplify each other in suppressing star formation.\footnote{Note that in our simulations $s_{\rm h} = 1 =
  \chi$ for $z\ge 9$ because there is no photo-heating at these
  redshifts.}
\par
To demonstrate that this amplification is indeed mutual, we write $s_{\rm wh}
\equiv s_{\rm w|h}s_{\rm h} \equiv s_{\rm h|w} s_{\rm w}$. This defines the
suppression factors $s_{\rm w|h} = \dot{\rho}_{*,\rm h}/ \dot{\rho}_{*,\rm wh}$ and
$s_{\rm h| w} = \dot{\rho}_{*,\rm w}/ \dot{\rho}_{*,\rm wh}$.  Thus, $s_{\rm w|h}$ is the
factor by which SN feedback suppresses the SFR in
simulations that include photo-heating and $s_{\rm h|w}$ is the factor by which
photo-heating suppresses the SFR in simulations that include
SN feedback. We have
\begin{equation}
\frac{s_{\rm w|h}}{s_{\rm w}} =
\frac{\dot{\rho}_{*,\rm h}/\dot{\rho}_{*,\rm wh}}{\dot{\rho}_{*}/ \dot{\rho}_{*,\rm w}} =
\frac{\dot{\rho}_{*,\rm w}/\dot{\rho}_{*,\rm wh}}{\dot{\rho}_{*}/ \dot{\rho}_{*,\rm h}}
=\frac{s_{\rm h|w}}{s_{\rm h}}.
\end{equation}
This shows explicitly that the amplification of the suppression of the SFR due to photo-heating in simulations that include SN 
feedback is equal to the amplification of the suppression of the SFR
due to SN feedback in simulations that include
photo-heating. It implies that we cannot determine whether photo-heating
amplifies the effect of SN feedback or vice versa.
\par
\begin{figure}
  \includegraphics[width=0.49\textwidth]{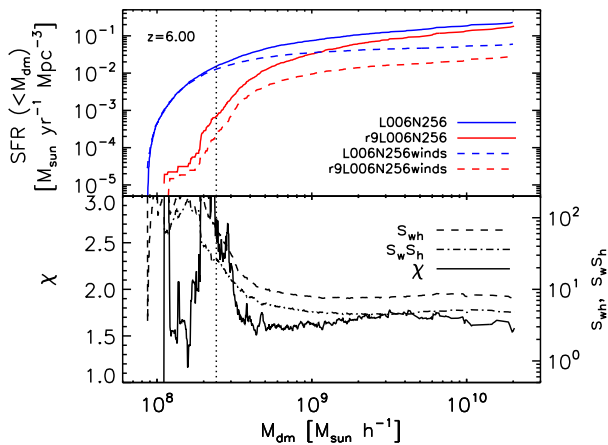}
  \caption{Top panel: cumulative SFRs as function of dark matter halo
    mass at $z = 6$ for
    our set of reference simulations. Photo-heating and SN feedback
    suppress the SFR over roughly complementary mass ranges. Bottom panel: similar to the
    bottom panel of Fig.~\ref{Fig:1}, but now for the cumulative SFRs
    shown in the top panel. The 
    vertical dotted lines indicate the dark matter mass corresponding to a virial
    temperature $T_{\rm vir} = 10^4 \K$.}
  \label{Fig:2}
\end{figure}

 
Fig.~\ref{Fig:2} shows, for our set of reference simulations, the dependence
of the suppression of the SFR due to photo-heating and/or SN feedback
on halo mass. The top panel shows the cumulative SFR at $z
= 6$ in dark matter halos less massive than $M_{\rm dm}$. The bottom
panel shows, similar to the bottom panel of Fig.~\ref{Fig:1}, the 
feedback amplification factor $\chi$ obtained from analogous definitions of
the suppression factors $s_{\rm h}, s_{\rm \rm w}$ and $s_{\rm \rm wh}$ applied to the
cumulative SFR at $z = 6$ shown in the top panel. The vertical
dotted lines indicate the dark matter mass corresponding to a virial
temperature $T_{\rm vir} = 10^{4}\K$.
\par
While photo-heating strongly decreases the SFR in halos with virial
temperatures $T_{\rm vir} \lesssim 10^{4}\K$, the inclusion of SN
feedback leads to a strong decrease in the SFR in halos with $T_{\rm vir} \gtrsim 10^{4}\K$. Radiative and kinetic feedback
thus act mostly over complementary mass ranges.  This dichotomy arises
because photo-evaporation mainly affects the gas fractions of halos with
masses that correspond to virial temperatures that are of order of or smaller
than the thermal temperature to which the gas is photo-heated, $T_{\rm vir}
\lesssim 10^{4}\K$. In contrast, star formation and the associated SN
feedback become only efficient for halos with masses corresponding to $T_{\rm vir} \gtrsim 10^{4}\K$, because our simulations do not
include radiative cooling from metals and molecules (and because of our limited resolution), which could
potentially enable star formation in halos of much smaller virial temperatures. 
\par
\begin{figure}
  \includegraphics[width=0.49\textwidth]{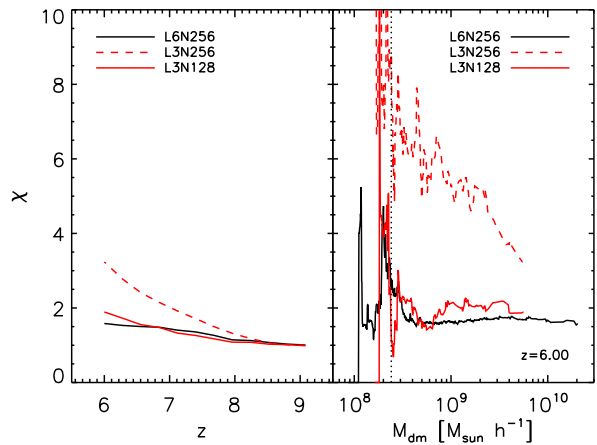}
  \caption{Evolution (left-hand panel) and mass-dependence (at $z =  6$;
    right-hand panel) of the feedback amplification factor $\chi$ for
    different choices of the box size and resolution. The vertical, dotted line in the right panel
    indicates the dark matter mass corresponding to a virial
    temperature $T_{\rm vir} = 10^4 \K$. The feedback amplification factor is insensitive
    to the size of the simulation box but increases strongly with resolution.}
  \label{Fig:3}
\end{figure}

The left-hand (right-hand) panels of Fig.~\ref{Fig:3} show the
evolution (mass-dependence) of the feedback amplification factor
$\chi$ obtained from simulations for which we have varied the size of the
simulation box and/or the resolution. Note that the solid, black
curves in the left-hand and right-hand panels are identical to the solid, black
curves shown in the bottom panels of Figs.~\ref{Fig:1} and
\ref{Fig:2}, respectively.
 \par
Changing the size of the simulation box by a factor of two (at fixed
resolution; solid curves) has little effect.  On
the other hand, increasing the resolution by a factor of two (while keeping
the size of the simulation box fixed; red curves) significantly
increases $\chi$. The 
increase in $\chi$ with increasing resolution is likely due to both an increase in
the fraction of galaxies that are subject to photo-evaporation
and an increase in the SFR (and thus associated SN feedback) of all galaxies.
\par
\par
\par
\section{Discussion}
\label{Sec:Discussion}
Photo-heating from reionisation and supernova (SN) feedback are key processes that determine the 
star formation rate (SFR) in the high-redshift Universe. Using a set of cosmological SPH simulations that
include radiative cooling and star formation, we analysed the $z \ge 6$ star formation history in
the presence of photo-ionisation heating from a uniform UV background and/or
kinetic feedback from core-collapse SNe. The inclusion of
photo-heating and SN feedback both lead to a suppression of the SFR.
\par
We showed that the factor by which the SFR is suppressed due
to photo-heating is larger in the presence than in the absence of SN
feedback. We also showed that the factor by which the SFR is
suppressed due to the inclusion of SN feedback is larger in the
presence than in the absence of photo-heating. This mutual amplification of
SN feedback and reionisation heating is the central result of the
present work. 
\par
We caution the reader that our simulations have not
fully converged with respect to resolution and that we have ignored
some potentially important physical processes. This result will
therefore need to be confirmed and quantified with future 
simulations. In what follows we briefly discuss the most important
physical effects that our analysis ignored. 
\par
We computed photo-heating rates from a uniform UV background in the optically
thin limit. Our simulations therefore do not account for the self-shielding of
gas from ionising radiation. This may lower the fraction of the
gas that is 
photo-evaporated (e.g., \citealp{Kitayama:2000}; \citealp{Susa:2004}; \citealp{Dijkstra:2004}). 
\cite{Iliev:2005} (extending the work of
\citealp{Shapiro:2004}) have, however, pointed out that regions that are
initially self-shielded will eventually also be photo-evaporated: as
subsequent layers of gas are photo-evaporated, previously 
self-shielded regions become exposed to ionising radiation and are eventually
stripped away. Moreover, the evaporation of the initially self-shielded gas
may proceed at a speed comparable to the speed predicted by simulations that
compute photo-heating rates in the optically thin limit (Fig.~3 in \citealp{Iliev:2005}). 
Note also that because SN explosions decrease the gas density, the effects of
self-shielding may be less prominent in simulations that include this type of
feedback.  Clearly, the importance of self-shielding and the applicability of
the optically thin limit to the present problem need to be critically
assessed using cosmological radiation-hydrodynamical simulations.
\par
Although all of our simulations employ a sufficiently high resolution to
resolve all halos with virial temperatures $T_{\rm vir} \gtrsim 10^4 \K$ with at least 100
particles, none of them has the resolution to properly reproduce the
properties of the multi-phase medium associated with the star-forming regions
these halos host. The factors by which photo-heating and SN feedback
suppress the SFR are therefore not yet converged. We have,
however, demonstrated that the effect of mutual amplification of photo-heating
and SN feedback becomes only stronger with increasing resolution.
\par
Because we have assumed the presence of a photo-dissociating 
background, the formation of molecular hydrogen is suppressed in our simulations.  
In reality, the gas may contain 
a significant fraction of molecular hydrogen before reionisation (but see, e.g., 
\citealp{Haiman:1997}). Star formation and the associated kinetic
feedback would then already be efficient in halos with virial temperatures
much smaller than $10^4\K$ (e.g., \citealp{Tegmark:1997}). Because the gas fraction 
in these halos would be affected by both photo-heating and SN feedback,
we expect that the inclusion of molecular hydrogen would only strengthen our
main result. 
\par
We have also ignored the existence of atoms and ions
heavier than helium. SN explosions may, however, quickly enrich the
interstellar and intergalactic gas with metals (e.g., \citealp{Bromm:2003}), which would increase its ability
to cool. Metal enrichment thus provides a
positive feedback that may partially offset the negative kinetic feedback from SN explosions.
\par
None of the caveats we discussed seems, however, likely to invalidate
our main qualitative conclusion that photo-heating and SN explosions
amplify each other's effect on the cosmic SFR. Galaxy formation
models that treat the effects of photo-heating and SN explosions independently
of each other, like e.g. (semi-) analytic models, may therefore significantly underestimate the effect of
these feedback processes on the SFR. Because photo-heating and SN
feedback are important processes that affect the gas in low mass-halos
at all epochs, our findings may also have implications for the
understanding of the properties of low-redshift galaxies, e.g. in the
context of the missing satellite problem (for a recent discussion see,
e.g., \citealp{Koposov:2009}).
\par

\section*{Acknowledgments} 
We are grateful to Volker
  Springel for letting us use his implementation of a Friends-of-Friends halo
  finder and we thank Claudio Dalla Vecchia for his implementation of additional
  variables.
Some of the simulations were run on the Cosmology Machine at
the Institute for Computational Cosmology in Durham as part of the
Virgo Consortium research programme. 
This work was supported by Marie Curie Excellence Grant
MEXT-CT-2004-014112.

\bsp
\label{lastpage}
\end{document}